\documentclass[twocolumn,showpacs,preprintnumbers,amsmath,amssymb,pre]{revtex4}

\usepackage{graphicx}      
\usepackage{dcolumn}       
\usepackage{psfig}

\begin{document}
\title{Thermo-statistical description of gas mixtures from space partitions}
\author{R. D. Rohrmann$^1$ and J. Zorec$^2$}
\affiliation{
$^1$ Observatorio Astron\'omico, Universidad Nacional de C\'ordoba, 
Laprida 854, X5000BGR C\'ordoba, Argentina\\
$^2$ Institut d'Astrophysique of Paris, UMR 7095 CNRS, Universit\'e 
Pierre \& Marie Curie, 98bis boulevard Arago, 75014 Paris, France
\\Electronic addresses: rohr@oac.uncor.edu, zorec@iap.fr
}

\date{\today}
\begin{abstract}
 The new mathematical framework based on the free energy of pure classical
fluids presented in [R. D. Rohrmann, Physica A {\bf 347}, 221 (2005)] 
is extended to multi-component systems to 
determine thermodynamic and structural properties of chemically complex
fluids. Presently, the theory focuses on $D$-dimensional mixtures in the
low-density limit (packing factor $\eta < 0.01$). 
The formalism combines the free-energy minimization
technique with space partitions that assign an available volume $v$ to each
particle. $v$ is related to the closeness of the nearest neighbor and provides
an useful tool to evaluate the perturbations experimented by particles in a 
fluid. The theory shows a close relationship between statistical geometry and
statistical mechanics. New, unconventional thermodynamic variables and 
mathematical identities are derived as a result of the space division. 
Thermodynamic potentials $\mu_{il}$, conjugate variable of the populations 
$N_{il}$ of particles class $i$ with the nearest neighbors of class $l$ are 
defined and their relationships with the usual chemical potentials $\mu_i$ are 
established. Systems of hard spheres are treated as illustrative examples and
their thermodynamics functions are derived analytically. The low-density 
expressions obtained agree nicely with those of scaled-particle theory and 
Percus-Yevick approximation. Several pair distribution functions are 
introduced and evaluated. Analytical expressions are also presented for hard 
spheres with attractive forces due to K\^ac-tails and square-well potentials.
Finally, we derive general chemical equilibrium conditions.
\end{abstract}

\maketitle
\section{ Introduction} \label{s.intr}

 The detailed knowledge of atomic and molecular po\-pu\-la\-tions is essential
to analyze and evaluate accurately thermodynamic properties and chemical 
processes in fluids as well as their monochromatic opacities. In particular,
they are required in studies of stellar interiors, atmospheres and 
circumstellar envelopes. Gas densities in stellar atmospheres and 
circumstellar structures are low, they are hardly higher than 0.001 
g\,cm$^{-3}$. Nevertheless, in these cases it is necessary to deal with a high 
variety of particles and electronic configurations and therefore, with a 
series of physical processes to calculate the radiation emitted by the star  
accurately \cite{kurucz,hubeny}.\par
 Among the most difficult challenges in determining gas models is to obtain
self-consistently atomic densities and opacity absorption coefficients by 
taking into account the non-ideal effects. The source of non-ideal effects is
the presence of Coulombic interactions between charged particles as well as
due to the short-ranged interactions between neutral-neutral and 
neutral-charged particles. Due to its relevance and complexity, the 
elaboration of gas models from mechanical statistical approaches is still
an active area of research  today \cite{fontaine,magni,robnik,mihalas,rogers,
saumon,potekhin,lisal,juranek,wini}.\par
 The present study is motivated by the difficulties of predicting 
quantitatively the optical properties of stellar atmospheres. In typical
gas models the opacity data are obtained through a complimentary calculation 
once the equations of state and mean abundances of bound states are obtained.
This procedure can produce, however, internal inconsistencies in a model, 
mainly when the abundances of atoms are calculated using internal energies 
that are different from those assumed to obtain the radiative transitions 
\cite{paper1,rohrmann}. This can lead to serious disagreements between 
predictions and observations or experiments (e.g., the unphysical Lyman 
opacity \cite{bergeron}). Therefore, it is highly desirable to develop a 
theoretical formalism that treats atoms subjected to different perturbations 
by considering them in separate groups 
and evaluate accordingly the populations and opacities in a consistent way. 
Such a gas model has then to consider explicitly not only all atomic 
excited states but also do they need to discriminate different perturbation 
states with the use of an appropriate particle-state variable. Our work is 
then oriented to develop a model with these characteristics.\par
 With the aim of establishing a detailed and accurate equation of state for
fluids, in a previous work we have developed a gas statistical formalism which 
combines free energy minimization methods with space partitions \cite{paper1}. 
The novel feature of this formalism, as compared to other more standard
and well-established statistical-mechanical theories of the fluid/liquid
state, is the attempt of keeping trace, in a thermodynamically self-consistent 
way, of a more refined structural information on the local environment of a 
given particle through an extra parameter, the ``available volume'', which is 
determined by the distance of the reference particle to its closest neighbor. 
For noninteracting particles, the available volume ``$v$'' is equivalent to
the spherical volume whose radius is the distance between the respective
centers of the particle and its nearest neighbor. This volume is considerably
smaller for particles with repulsive interactions. The theory \cite{paper1}
was formulated 
for one-component dilute fluids, where the Helmholtz free energy is written in
terms of the occupation number distribution $N_v$ of the $v$-variable and 
where the particle interactions are introduced using pair potentials. The 
result thus obtained represents a unified treatment of thermodynamics and the
structure of fluids. The formulation was applied to hydrogen atoms in an 
electrically neutral medium of charged particles encompassing ions and 
electrons. The atomic populations were derived by establishing groups of atoms 
having different plasma perturbations measured according to the size of their
available space volumes $v$. Although the results obtained are only 
illustrative of its possible performances, this theory represents the first 
attempt of evaluating in detail the populations by combining consistently 
the plasma interaction effects.\par
 Approaches based on space partitions were attempted in the past with the 
so-called ``cell" theories to obtain equations of state for fluids. However,
the character of these space divisions is quite different as compared to that
employed in \cite{paper1} and in the present paper. The cell models 
\cite{hcb,hill,kirk} were originally developed by Eyring and Hirschfelder 
\cite{eyring} and by Lennard-Jones and Devonshire \cite{lennard}. They were 
inspired by the lattice-like configurations acquired by some systems at high
densities. In such theories the fluid volume is divided into an imaginary 
lattice of cells. Each particle is then confined into a cell, which implies 
that a space partition is assumed beforehand, instead of being deduced from 
the model proper. Besides, by ignoring the interchange of particles between 
cells, an error is introduced in the system-entropy calculation, which is 
usually corrected using an arbitrary ``communal entropy'' factor.\par
 The perspective of the theory presented in \cite{paper1} is new, as it
determines a space division by assigning the $v$-volume to the particles. This 
is achieved by minimizing the system free energy under the condition that the
sum of all $v$-volumes be equivalent to the entire volume of the fluid. In 
this procedure, no particle confinement is established, because the assignment 
of an individual volume to each particle does not introduce any ownership 
connection between space regions and particles, i.e., $v$ does not carry any 
information on the particle coordinates. Therefore, the formulation presented
in \cite{paper1} is free from difficulties that are inherent to the cell 
theories, i.e. entropy corrections, and it can provide new results, such as
the combined thermodynamical and structural information of a fluid, which 
cannot be derived from the cell models.\par
 In the present paper we generalize the formalism to $D$-dimensional
multi-component fluids that was initially developed in \cite{paper1} only
for pure or single-component substances. It aims at giving particularly simple 
analytic forms to provide a unified derivation of thermodynamic and geometric 
properties that characterize them. The formulation of the theory is general,
but attention is focused on its application to hard sphere models, which 
represent the repulsive interactions among atoms and geometrically simple 
molecules \cite{barker}. We also include the analysis of hard spheres with 
weak long-range attraction (K\^ac type) and hard spheres with square-well 
potential. The approximations discussed here should be suitable for more 
realistic short-ranged potentials, like those governing many atomic and 
colloidal fluids. 
 Our introduction of space partitions in statistical 
thermodynamics marks a considerable progress in the description of gases with
respect to previous theories, since it gives the possibility of obtaining 
thermodynamic and structural information on fluids simultaneously and
self-consistently. The present extension to multicomponent fluids enables us 
to treat arbitrarily complex chemical mixtures by including systematically the
inter-particle interaction potentials. The theory can have many applications,
not only in Astrophysics, but also in other fields such as the analysis of 
colloidal and protein solutions in biochemistry.\par
 In the following sections, we first introduce a description of states of a
gas mixture based on space partitions. General expressions for the Helmholtz 
free energy are presented in Sect. \ref{s.Helm} and the thermodynamic 
equilibrium states are derived in Sect. \ref{s.equ_sta}. We analyze there some
non-conventional thermodynamic variables that are generated by the theory 
and derive general expressions to calculate chemical potentials. Then, 
applications of the formalism to a few multi-component systems based on hard 
particles with and without attractive forces are given in Sect. \ref{s.hard}. 
The results obtained are compared with those from Percus-Yevick and scale 
particle theories. In Sect. \ref{s.chem} we extend the theory to include 
chemical reactions in the gas. 
The conclusions are presented in Sect. \ref{s.concl}.\par

\section{Gas states}

The approach developed here is inspired by the variety of spatial patterns 
shown by fluids where the particles have different types of interactions. 
The main assumption that will be made is that the spatial structure of a gas 
in equilibrium can be described by an appropriate division of the space among 
the particles. The study of single-component fluids at low densities 
\cite{paper1} has shown that the volume $v$ assigned to a particle depends 
on the interactions the particle has with the closest neighbor. 
In Sect. \ref{s.hard} we give a formal definition of the closest neighbor. 
According to the mentioned result, the generalization of the formalism 
to multi-component fluid systems (mixtures), requires that the particle 
population $N_i$ of each chemical species $i$ be divided into sub-groups 
or sub-classes in accordance with the chemical species of the 
nearest neighbor, i.e.
\begin{equation} \label{e.Ni}
N_i = N_{i1} + N_{i2} + N_{i3} + ...,
\end{equation}
\noindent where the sub-indices 1,2,3... specify the various chemical species 
of the closest neighbors. We consider that for any reference particle in 
whatever configuration of $N$ particles, there is `one and only one' $j$th 
neighbor, with $1\le j \le N-1$. 
This means that if there is degeneracy where two 
particles can be equidistant from the reference, one of them will be 
designated as the $j$th neighbor arbitrarily, while the other will be the 
$(j+1)$th neighbor. In what follows we shall use the sub-index $ilv$ for a 
particle of class $i$ that has individual available volume $v$, whose nearest 
neighbor is of class $l$. The state of the gas will then be described by the
following collections:
\begin{equation} 
\label{e.state}
\left[\begin{array}{ccc}
N_{11v} & N_{12v} & \ldots \\
N_{21v} & N_{22v} & \ldots \\
\vdots  & \vdots  & \ddots \\
\end{array} \right] \hspace{.3in} (0\le v\le V),
\end{equation}
\noindent where $N_{ilv}dv$ is the number of particles $ilv$ with available
individual volumes ranging from $v$ to $v+dv$. The physically acceptable states
in (\ref{e.state}) must obey the particle number $N_{il}$ and gas volume $V$ 
conservation conditions given~by
\begin{equation} \label{e.Nil}
N_{il} = \int_0^V\!N_{ilv}dv,
\end{equation}
\begin{equation} \label{e.Vcons}
V = \sum_{il}\!\int_0^V\!\!vN_{ilv}dv.
\end{equation}
Thus, the individual volume $v$ assigned to each particle acts as an
attribute inherent to each particle and the sum of $v$-spaces is equivalent to
the entire volume of the gas. This space partition does not involve any
knowledge of particle positions, since the assignment of a $v$-volume
to each particle does not carry any information on the particle
coordinates.\par

\section{Helmholtz free energy} \label{s.Helm}

 To evaluate the free energy of the gas we use a composition of translation-,
configuration- and interaction-dependent terms:
\begin{equation} \label{e.F3}
F = F_{trans} + F_{conf} + U_{int},
\end{equation}
\noindent which is suitable for a large variety of fluids \cite{hansen}. The 
first two contributions to $F$ can be written respectively as:
\begin{equation} \label{e.Ftrans}
F_{trans} = \sum_{il}\!\int_0^V\!\frac{N_{ilv}}\beta\ln\left(\frac 
{\lambda_i^D}V\right)dv,
\end{equation}
\begin{equation} \label{e.Fconf}
F_{conf} = \sum_{il}\!\int_0^V\!\frac{N_{ilv}}\beta 
\ln\left(\frac{N_{ilv}V}{N_i}\right)dv,
\end{equation}
\noindent where $\beta=(kT)^{-1}$, $k$ is the Boltzmann constant, $T$ is the
temperature and $\lambda_i$ is the thermal wavelength of particles of class 
$i$. $F_{conf}$ was obtained following the steps described in \cite{paper1}, 
except that in the present case was done for a gas-mixture. For the 
multi-component system we have then that the total number of ways $W$ to 
generate a set $\{N_{ilv}\}$ of occupation numbers from the individual 
complexions $W_i$ is $W=\prod_iW_i$, where: 
\begin{equation} \label{e.Wi}
W_i = N_i!{\prod_{lv}}\frac{(dv/V)^{N_{ilv}dv}}{(N_{ilv}dv)!}.
\end{equation} 
\noindent Therefore, according to \cite{paper1} the configuration entropy 
is:
\begin{equation}
S_{conf} = k\ln W = -k\sum_{il}\!\int_0^V N_{ilv}\ln\left(\frac{N_{ilv}V}
{N_i}\right)dv,
\end{equation}
\noindent which finally determines (\ref{e.Fconf}), since $F_{conf}=
-TS_{conf}$.\par
 The interaction can be generalized in the same way as follows
\begin{equation}
U_{int} = \sum_{il}\!\int_0^V\!N_{ilv}\!\left(\sum_j\frac{N_j\phi_{ilv,j}}{2V}
\right)dv,
\end{equation}
\noindent where the pair-interaction factors $\phi_{ilv,j}$ are:
\begin{equation} \label{e.phi}
\phi_{ilv,j} = \int_0^{V'}\!u_{ilv,j}(\omega)g_{ilv,j}(\omega)d\omega.
\end{equation}
\noindent In (\ref{e.phi}), $u_{ilv,j}$ is the interaction potential between
a particle of class $ilv$ and another of class $j$. The distribution function
of particles $j$ with respect to the reference particle $ilv$ is written as
$g_{ilv,j}$. We note that the integration variable $\omega$ in Eq. (\ref{e.phi})
is a spherical volume centered at the particle of class $i$ and that $V'$ 
denotes the 
minimum $\omega-$volume able to enclose the entire volume of the mixture. The 
total free energy of the gas-mixture is then:
\begin{equation} \label{e.F}
F\!=\! \sum_{il}\!\int_0^V\! \!N_{ilv}\left[\!\frac{1}{\beta}\ln
\left(\!\frac{N_{ilv}\lambda_i^D}{N_i}\!\right)\!+\!\sum_j\!\frac
{N_j\phi_{ilv,j}}{2V}\!\right]\!dv.
\end{equation}

\section{ Equilibrium States} \label{s.equ_sta}

 The equilibrium distributions $N_{ilv}$ are obtained by minimizing the free
energy (\ref{e.F}) and taking into account the conservation conditions given 
by Eqs. (\ref{e.Nil}) and (\ref{e.Vcons}). The minimizing procedure must take
into account that the variations $\delta N_{ilv}$ affect the total particle 
populations of each species according to the relation:
\begin{equation} \label{e.Nj}
N_j = \sum_m \int_0^V\!N_{jmv}dv.
\end{equation}
We thus obtain for the state of equilibrium:
\begin{equation} \label{e.Nv}
N_{ilv} = \frac{N_i}{\lambda_i^D} \exp \left[ -\beta \left( \gamma v 
- \mu_{il} + \epsilon_{ilv} + \epsilon_i \right) \right],
\end{equation}
\noindent where
\begin{equation} \label{e.Eilv}
\epsilon_{ilv} = \sum_j\frac{N_j\phi_{ilv,j}}{2V}
\end{equation}
\noindent and
\begin{equation} \label{e.Ei}
\epsilon_i = \sum_{jm}\frac 1{2V}\!\int_0^V\!N_{jmv}\phi_{jmv,i}dv.
\end{equation}
\noindent In Eq. (\ref{e.Nv}), $\mu_{il}$ and $\gamma$ are the Lagrange
multipliers corresponding to conditions (\ref{e.Nil}) and (\ref{e.Vcons})
respectively. Accordingly, the potentials $\mu_{il}$ determine the total 
population of particles $i$ and their redistribution into sub-sets as a 
function of the nearest neighbor class. The variable $\gamma$ 
determines the distribution of individual available volumes $v$.\par
 It is useful to note that the total interaction energy can also be written as
\begin{equation}
U_{int}=\sum_i N_i \epsilon_i = \sum_{il}\int_0^V\!N_{ilv}\epsilon_{ilv}dv,
\end{equation}
\noindent which enable us to see that $\epsilon_i$ is the average energy due 
to all interactions the particle of class $i$ has with the gas, 
while $\epsilon_{ilv}$ represents the interaction energy of the $ilv$
particle with the gas. Finally, using 
(\ref{e.Nv}) and (\ref{e.F}) we can write the equilibrium free energy for 
multi-component gases in the Eulerian form 
\begin{equation} \label{e.Fequil}
F = -\gamma V+\sum_{il}\mu_{il}N_{il} -U_{int}.
\end{equation}
  
\subsection{Lagrange parameters}\label{s.lagran7}

 Relation (\ref{a.ide}) obtained in (Sect. \ref{s.iden}) and (\ref{e.Fequil})
help us to show easily that $\mu_{il}$ behaves as a thermodynamic potential
associated to the $N_{il}$ population:
\begin{equation} \label{e.muil}
\mu_{il} = \left. \frac{\partial F}{\partial N_{il}}\right|_{T,V,N_{jm\ne 
il}}.
\end{equation}
 Similarly, making use of (\ref{e.Fequil}) and (\ref{a.ide2}) we find that
$\gamma$ relates to the pressure $P$ in the already known way for 
single-component fluids \cite{paper1},
\begin{equation} \label{e.pgu}
P = \gamma+\frac{U_{int}}{V},
\end{equation}
\noindent i.e., the gas pressure is the sum of $\gamma$ and the 
interaction-energy density.

\subsection{Chemical potentials}\label{s.chempot}

 The chemical potential corresponding to $i-$class particles is: 
\begin{equation} \label{e.mui}
\mu_i = \left. \frac{\partial F}{\partial N_i}\right|_{T,V,N_{j\ne i}}.
\end{equation}
\noindent The substitution of (\ref{e.Fequil}) into (\ref{e.mui}) and the use 
of relation (\ref{a.ide3}) lead to the following expression of $\mu_i$ in 
terms of potentials $\mu_{il}$,
\begin{equation} \label{e.mui2}
\mu_i = \sum_{lm} \mu_{lm} \left. \frac{\partial N_{lm}}{\partial N_i}
\right|_{T,V,N_{j\ne i}}.
\end{equation}
 In fluids with low enough density, it is expected that the splitting up of 
the $N_i$ populations into sub-groups $N_{il}$ [Eq. (\ref{e.Ni})] be 
determined by the mole fractions $x_l=N_l/N$ of species present in the gas.
This means that the population of particles $i$ having as closest neighbors 
the class $l$ particles is proportional to the total population of 
$i-$particles and to the abundance of $l-$class particles:
\begin{equation} \label{e.nilninl}
N_{il} = \frac{N_l N_i}{N} = x_lN_i.
\end{equation}
\noindent We can also say that the probability of a particle had an $l-$class
as the closest neighbor is given simply by the abundance of $l$ species. In 
this case (\ref{e.mui2}) becomes
\begin{equation} \label{e.mui3}
\mu_i = \sum_l(\mu_{il}+\mu_{li})x_l-\sum_{lm}\mu_{lm}x_lx_m.
\end{equation}
 It is easy to show that if $\mu_{il}$ is given by the sum of two terms $f_i+
g_l$, i.e., one term that depends on the class $i$ and the other on $l$, it 
will be $\mu_i=f_i+g_i$. This is what happens in a mixture of ideal gases, as
it is shown in what follows. Let us note that introducing (\ref{e.Nv}) into 
(\ref{e.Nil}) leads to the useful relation
\begin{equation} \label{e.nilmuil}
N_{il} = \lambda_i^{-D}N_i Z_{il}e^{\beta\mu_{il}},
\end{equation}
\noindent where
\begin{equation} \label{e.Zdef}
Z_{il}=\int_0^V\!\exp\left[-\beta\left(\gamma v+\epsilon_{ilv}+
\epsilon_i\right) 
\right]dv.
\end{equation}
\noindent On one hand, for non-interacting gases (hereafter identified with the 
superscript $^*$), we have
\begin{equation}
Z_{il}^*= (\beta\gamma)^{-1}=n^{-1},
\end{equation}
\noindent which is obtained from (\ref{e.Vcons}) and the estimation is done
for the thermodynamic limit $(N,V)\!\to\!\infty$ with $n\equiv N/V\!=$ 
constant. Therefore, for a mixing of gases it becomes
\begin{equation}
\mu_{il}^*= kT \ln \left( \frac{N_l\lambda_i^D}{V}\right),
\end{equation}
\noindent which through Eq. (\ref{e.mui3}) yields the following exact result 
\begin{equation}
\mu_i^*= kT\ln\left(\frac{N_i\lambda_i^D}{V}\right).
\end{equation}
 On the other hand, for interacting fluids, relation (\ref{e.Zdef}) gives
\begin{equation} \label{e.Zilg}
Z_{il} = (\beta\gamma)^{-1}q_{il}e^{-\beta\epsilon_i},
\end{equation}
\noindent with
\begin{equation} \label{e.qil}
q_{il}=\int_0^\infty\!e^{-(x+\beta \epsilon_{ilx})}dx,
\end{equation}
\noindent where $x=\beta\gamma v$. In general, if interactions in the 
gases are fairly low, factors (\ref{e.qil}) will be $q_{il}\approx1$. In fact,
$q_{il}\!\to\!1$ if $\epsilon_{ilv}\!\to\!0$. Using the condition (\ref{e.nilninl}) 
one obtains
\begin{equation} \label{e.milg}
\frac{\mu_{il}^{ex}}{kT} = \ln\left(\frac{\beta\gamma}n\right)+ 
\frac{\epsilon_i}{kT}-\ln q_{il},
\end{equation}
\begin{equation} \label{e.mui_a}
\frac{\mu_i^{ex}}{kT}= \ln\left(\frac{\beta\gamma }n\right)+ 
\frac{\epsilon_i}{kT}-\ln Q_i,
\end{equation}
\noindent in which
\begin{equation}
\ln Q_i = \sum_lx_l\left[\ln(q_{il}q_{li})-\sum_m x_m\ln q_{im}\right]. 
\end{equation}
 The excess chemical potentials $\mu_{il}^{ex}=$ $\mu_{il}-\mu_{il}^*$ and 
$\mu_i^{ex}=$$\mu_i-\mu_i^*$ are formed by three contributions with specific 
characteristics. The first right-hand term of equations (\ref{e.milg}) and 
(\ref{e.mui_a}) represents the deviation of the division of space among 
particles with respect to that in a perfect gas, i.e. $\beta\gamma\ne n$. This 
effect is the same for all chemical potentials. The second right-hand term in
(\ref{e.milg}) and (\ref{e.mui_a}) gives the contribution due to the potential
energy acquired on average by an $i-$class particle due to its interaction
with all the remaining particles in the gas. The third term corresponds to 
interactions that affect the distribution of individual spaces $v$, as can be
seen from definition of $q_{il}$ in (\ref{e.qil}). This last contribution can
be significant even for interactions that do not carry energy to the gas, as 
it is for infinitely strong repulsive forces (see Sect. \ref{s.hard}).\par
 By virtue of Eq. (\ref{e.mui_a}), we can obtain a similar expression to 
(\ref{e.nilmuil}) for the total population of a given class
\begin{equation} \label{e.Nig}
N_i = \lambda_i^{-D}NZ_ie^{\beta\mu_i},
\end{equation}
\noindent where
\begin{equation} \label{e.Zig}
Z_i = (\beta \gamma)^{-1}Q_ie^{-\beta\epsilon_i}.
\end{equation}

\section{Hard particles} \label{s.hard}

 A mixture of hard spheres in $D$ dimensions is characterized by interaction 
potentials described by:
\begin{equation} \label{e.uv_hs}
u_{ilv,j}(\omega) = \left\{ 
\begin{array}{rcl}
           \infty;  & &\omega \le a_{ij} \\  
                0;  & &\omega > a_{ij}   \\
\end{array}
\right.,
\end{equation}
\noindent with $a_{ij}=2^D\sigma_{ij}$, where $\sigma_{ij}$ is the volume of
a $D$-dimensional sphere with diameter $(d_i + d_j)/2$, 
being $d_i$ and $d_j$ the respective
diameters of particles of classes $i$ and $j$. The interaction does not depend 
neither on the $l-$class neighbor nor on the available volume. To identify 
the available volume $v$ of a given particle we adopt as reference the results 
found in \cite{paper1} for single-component fluids. We adopt then as ansatz 
that for low particle densities the available volume is the ideal volume $v^*$
(i.e., $v$ of non-interacting particles, see Sect. \ref{s.intr})
reduced by an amount $a_{il}^*$ that depends on the repulsion between the
particle and its {\it neighbor with the shortest contact distance} (hereafter
$nscd$):
\begin{equation} \label{e.vid}
v = v^*-a_{il}^*.
\end{equation}
 Other interpretations than (\ref{e.vid}) are also possible. However, the 
results that are obtained in the present section strongly support the choice 
given in (\ref{e.vid}). As a consequence of this interpretation, a particle of 
class $j\ne l$ with $a_{ij}<a_{il}$ can have its center at a shorter distance 
from the center of particle $i$ than the particle $l$.\par
 The identification of a volume per particle enable us to specify the form of
the distribution function of pairs (hereafter $pdf$) $g_{ilv,j}(\omega)$ for 
low densities. Notice that $g_{ilv,j}(\omega)d\omega$ gives the probability of 
finding a particle of class $j$ located between the surfaces of spheres 
$\omega$ and $\omega+d\omega$ both centered in a particle $ilv$. According to
the results obtained in \cite{paper1} for single-component fluids, we deduce
that the $pdf$ with $j=l$ has a contribution due to the $nscd$ given by a 
Dirac delta function $\delta(x)$ evaluated on the $nscd$ center and divided by 
the density of $l-$class particles. In fluids of low density, the contribution 
to the $pdf$ due to the remaining particles $j$ can be approximated with a
Heaviside step function $\Theta(x)$ [$\Theta(x)=1$ for $x\ge 0$ and 
$\Theta(x)=0$ when $x< 0$], which implies that the position of these 
particles is uncorrelated with that of the reference $ilv$ one. Thus,
\begin{equation}\label{e.ghard}
g_{ilv,j}(\omega)\!=\!\frac{\delta_{jl}}{n_l}\delta(\omega-v-a_{il}^*)\!+\! 
\Theta(\omega\!-\!v\!-\!a_{il}^*\!+\!a_{il}\!-\!a_{ij})\!.
\end{equation}
\noindent The terms in the argument of the step function warrant that the
$nscd$ is a $l-$class particle. The Kronecker delta symbol $\delta_{jl}$ 
ensures that the contribution due to $nscd$ is taken into account only if
$j=l$. Introducing Eqs. (\ref{e.uv_hs}) and (\ref{e.ghard}) into Eq. 
(\ref{e.phi}) we obtain
\begin{equation} \label{e.phil}
\phi_{ilv,l} = \left\{ 
\begin{array}{rcl}
           \infty; & &  v \le b_{il} \\  
                0; & & v > b_{il}    \\
\end{array}
\right. ,
\end{equation}
\noindent and 
\begin{equation}  \label{e.phij}
\phi_{ilv,j \ne l} \equiv 0 \,,
\end{equation}
\noindent where
\begin{equation} \label{e.baa}
b_{il} = a_{il} - a_{il}^* .
\end{equation}
 It follows then that
\begin{equation} 
\epsilon_{ilv} = \left\{ 
\begin{array}{rcl}
           \infty;    & & v \le b_{il} \\  
                0;    & & v > b_{il}   \\
\end{array}
\right.,
\end{equation}
\noindent while $\epsilon_i=0$ $\forall i$. In the thermodynamic limit, Eqs.
(\ref{e.Nil}) and (\ref{e.Vcons}) yield
\begin{equation} \label{e.a_hs}
\mu_{il} = kT\left[\frac{b_{il}n}{1-bn}+
\ln\left(\frac{n_{il}\lambda_i^D}{n_i}\frac{n}{1-bn}\right)\right],
\end{equation}
\begin{equation}\label{e.nv}
n_{ilv} = \Theta(v-b_{il}) n_{il} \beta \gamma \exp\left[-\beta\gamma 
\left(v-b_{il}\right)\right],
\end{equation}
\begin{equation}\label{e.P}
\beta\gamma = \frac{n}{1-bn},
\end{equation}
\noindent where we have written
\begin{equation} \label{e.bn}
bn = \sum_{il}b_{il}n_{il}.
\end{equation}
 As it is expected, the gas does not store interaction energy, which implies
that $U_{int}=0$ and thus $\gamma=P$. From the comparison of Eq. (\ref{e.P}) 
with the corresponding in the Virial expansion for mixtures of hard spheres
\cite{hcb}, enable us to make the following identifications
\begin{equation}\label{e.bij}
b_{il} = 2^{D-1}\sigma_{il},
\end{equation}
\begin{equation}\label{e.mol}
n_{il} = x_ln_i.
\end{equation}
\begin{figure}
 \scalebox{0.40}{\includegraphics{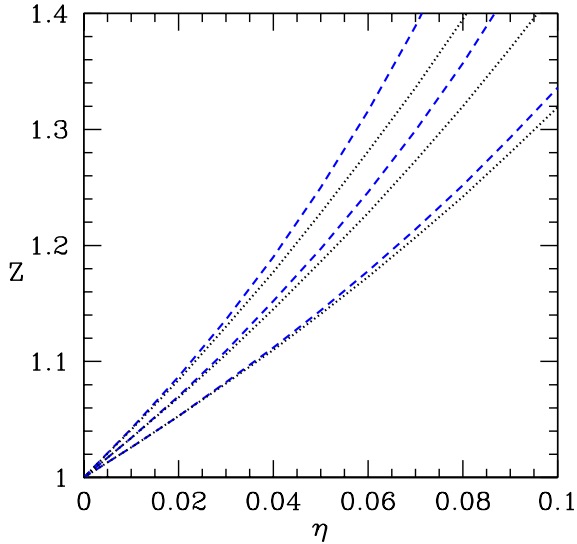}}
\caption{\label{f.zbin} (Color online). 
The compressibility $Z$ factor calculated with Eq. 
(\ref{e.P}) (dashed lines) and Percus-Yevick results \cite{lebowitz} (dotted
lines) as a function of the packing fraction $\eta$ for $3D$ binary mixtures
with $x_1=0.5$ and $d_2/d_1=0.01$, $0.4$, and $1.0$ (from bottom to top).} 
\end{figure}
 Relation (\ref{e.bij}) is nothing but an extension of previous results 
found for single-component fluids, while (\ref{e.mol}) confirms the
relation already established in  Eq. (\ref{e.nilninl}).
Since $\sigma_{il}=\sigma_{li}$ the quantities $a_{il}$, $b_{il}$, and 
$a^*_{il}$ are symmetric in the indices $il$. In particular, $a^*_{il}=$
$a^*_{li}=$$ 2^{D-1}\sigma_{il}$, so that the volume of a hard sphere is 
reduced with respect to the ideal value $v^*$ by an amount that does not 
depend on which of both, the reference or the $nscd$ particle, is the largest.
Nevetheless, as it is expected, a big $nscd$ reduces more the space $v$ of
a given particle than a small $nscd$.\par
 On the other hand, from Eqs. (\ref{e.mui3}) and (\ref{e.a_hs}) we obtain the 
excess chemical potential for hard spheres of species $i$ that is given by
\begin{equation} \label{e.exmu_hs}
\frac{\mu_i^{ex}}{kT} = -\ln(1-bn)+\frac{2\sum_lb_{il}n_l-bn}{1-bn},
\end{equation}
\noindent which in particular for $D=1$ reduces to
\begin{equation}  \label{e.exmu_rods}
\frac{\mu_i^{ex}}{kT}=-\ln\left(1-\sum_l\sigma_{ll}n_l\right)+\frac{\sigma
_{ii}n}{1-\sum_l\sigma_{ll}n_l}.
\end{equation}
\begin{figure}
 \scalebox{0.40}{\includegraphics{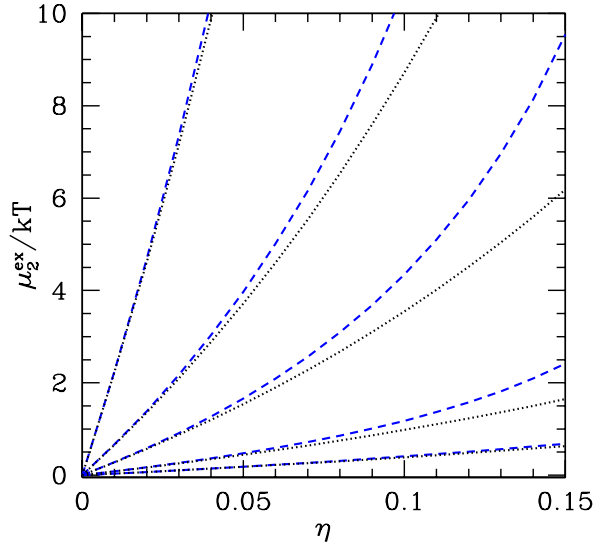}}
\caption{\label{f.Reiss} (Color online). 
Excess chemical potentials of a diluted solution 
(here species 2) calculated with Eq. (\ref{e.exmu_hs}) (dashed lines) and those
from the scale particle theory \cite{reiss} (dotted lines) as a function of 
the packing fraction $\eta$ for $3D$ binary mixtures with $x_2=10^{-5}$ and 
$d_2/d_1=0.5$, $1.0$, $2.0$, $3.0$, and $5.0$ (from bottom to top).} 
\end{figure}
\begin{figure}
 \scalebox{0.40}{\includegraphics{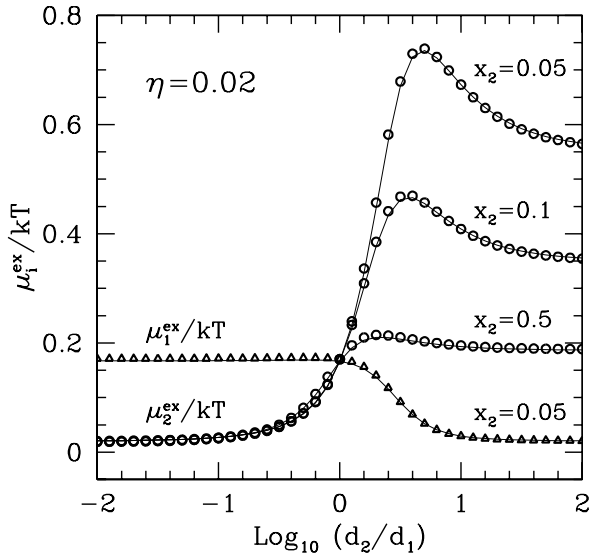}}
\caption{\label{f.d2d1} Excess chemical potentials obtained with Eq. 
(\ref{e.exmu_hs}) (symbols) and Percus-Yevick theory results \cite{salacuse} 
(lines) as a function of the diameter ratio $d_2/d_1$ for $3D$ binary mixtures
with $\eta=0.02$ and $x_2=0.05$, $0.1$, and $0.5$.} 
\end{figure}
 Eq. (\ref{e.P}) is exact for $D\!=\!1$ \cite{salsburg,stell} and it gives 
only an approximate solution for $D\!>\!1$. However, it gives the correct 
asymptotic behavior for $n\!\to\!0$. This is shown explicitly in 
Fig.~\ref{f.zbin}, where the compressibility factor $Z=P/(nkT)$ is calculated
with Eq. (\ref{e.P}) and compared with the Percus-Yevick (PY) approximation 
\cite{lebowitz} for a 3--dimensional binary mixture, as a function of the 
packing fraction $\eta=$ $\sum_in_i\sigma_{ii}$ which gives the fraction of 
the total volume occupied by the hard particles. 
 Since we compare our results with those of the PY theory, one of the
most successful theoretical approaches in studies of fluids, let us recall
that it is based on an approximate integral equation of the radial
distribution function from which the thermodynamics of hard-body systems 
follows as a by product. In few cases, it can be solved analytically,
e.g. mixture of hard spheres, and the solutions obtained are very close to
those issued from computer simulation \cite{hansen}.\par
 Likewise, Eq.~(\ref{e.exmu_rods}) is exact for hard rods \cite{salacuse}. 
Eq.~(\ref{e.exmu_hs}) provides the correct low-density behavior of the 
excess chemical potential for $D>1$. This last is illustrated in
Figs.~\ref{f.Reiss} and \ref{f.d2d1}, which show the comparisons of our
results obtained for $D=3$ with those issued from the scale particle theory
\cite{reiss}, which is correct for a solute infinitely diluted
and with the PY solutions found by Lebowitz \cite{lebowitz},
respectively. It is important to note that all expressions derived for 
mixtures of hard-spheres are exact up to the second term in the virial
expansion. Finally, we can conclude that the achievement of agreements shown
in Figs. \ref{f.zbin}, \ref{f.Reiss} and \ref{f.d2d1} confirm the 
appropriateness of the closure relations (\ref{e.nilninl}) and (\ref{e.vid}) 
adopted, where $a^*_{il}=$ $2^{D-1}\sigma_{il}$.
Since our work aims at giving a detailed representation of complex 
gases at low densities ($\eta<0.01$), the results we have obtained can be
considered as very satisfactory.

\subsection{Pair-distribution functions} \label{s.gij}
 
 From detailed $g_{ilv,j}$ functions it is possible to calculated the averaged
$pdf$ with the occupational numbers $n_{ilv}$ (see \cite{paper1}). For example,
the probability of finding a $j$ particle between the spherical surfaces 
$\omega$ and $\omega+d\omega$ centered in a $i$ particle, regardless 
of its available volume, whose $nscd$ is an $l$ particle, defines a
{\em conditional} pair distribution function $cpdf$ which is given by
\begin{equation} \label{e.gmel}
g_{il,j}(\omega) = \frac{1}{n_{il}}\int_0^\infty\!g_{ilv,j}(\omega)n_{ilv}dv.
\end{equation} 
The traditional $g_{ij}$ can be computed from the summation over all species
of $cpdf$ appropriately weighted by the probability of finding a $nlcd$ of a 
given class,
\begin{equation} \label{e.gme}
g_{ij}(\omega) = \sum_lx_lg_{il,j}(\omega).
\end{equation}
For mixtures of hard-spheres, Eqs. (\ref{e.gmel}) and (\ref{e.gme}) give 
respectively,
\begin{equation} \label{e.gl}
g_{il,j}(\omega)\!=\!\Theta(\omega-a_{ij})\!\left[\!1\!+\!\left(\delta_{jl} 
\frac{\beta\gamma}{n_l}\!-1\!\right)\!e^{-\beta\gamma(\omega-a_{ij})}\! 
\right] ,
\end{equation}
\begin{equation} \label{e.gij}
g_{ij}(\omega)\!=\!\Theta(\omega-a_{ij})\!\left\{\!1\!+\!\frac{bn}{1-bn}\! 
\exp\left[\!-\frac{n(\omega-a_{ij})}{1-bn}\!\right]\!\right\}.
\end{equation}
Eqs. (\ref{e.gl}) and (\ref{e.gij}) are approximations suitable for low 
densities, because the function $g_{ilv,j}$ in Eq. (\ref{e.ghard}) takes into
account explicitly only the correlations between the reference particle and 
its $nscd$. However, they contain the most important deviations with respect 
to the solutions for perfect gases.\par
\begin{figure}
\scalebox{0.43}{\includegraphics{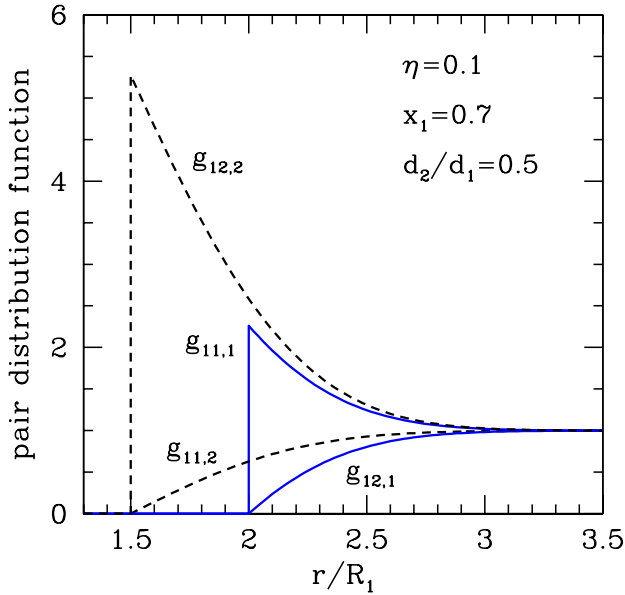}}
\caption{\label{f.gilj} (Color online). 
Conditional p.d.f. from Eq. (\ref{e.gl}) as a function 
of the normalized radial distance $r/R_1$ ($R_1=\frac{d_1}2$ is the radius of 
a particle of species 1) for a $3D$ binary mixture with $\eta=0.1$, $x_1=0.7$, 
and $d_2/d_1=0.5$.} 
\end{figure}

 To our knowledge, this is the first time that the $cpdf$ are suggested and 
calculated for mixtures of hard spheres. Figure~\ref{f.gilj} shows some $cpdf$
for a binary mixture with $\eta\!=\!0.1$ and $x_1\!=\!0.7$. In the chosen 
example, particles of class 1 have a diameter twice as large as those of class 
2. Functions $g_{il,j}$ become zero for distances $r$ between particles $i$ 
and $j$ shorter than the contact distances, as it happens for the conventional 
$pdf$ $g_{ij}$. Nevertheless, contrary to functions $g_{ij}$, the behavior of 
$cpdf$ at moderate distances can be strongly affected due to the information 
on the $nscd$ class carried by the reference particle. Thus, up to the present 
level of approximations, the function $g_{12,2}$ grows above 1 for small 
values of $r$ ($r\!>\!1.5R_1$, $r\!\to\!1.5 R_1$, where $R_1$ is the radius of class 1
spheres). This is due to the fact that the probability of finding a particle
of class 2 at a distance $r\approx1.5R_1$ from another of class 1 augments if 
it happens that this last has a $nscd$ of class 2. This augmentation concerns
the non-conditioned probability of finding a class 2 particle and so, it is 
higher the lower is the abundance of the considered species. Conversely, for 
the same range of distances, the function $g_{11,2}$ decreases because the 
possibility of $nscd$ be of class 2 is removed.\par
\begin{figure}
\scalebox{0.43}{\includegraphics{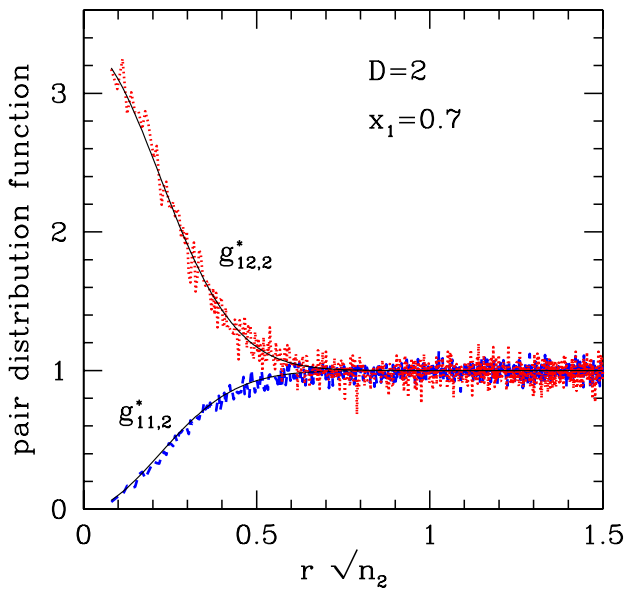}}
\caption{\label{f.gide} (Color online). 
Conditional p.d.f. computed with Eq. (\ref{e.gl}) 
(full lines) as a function of the dimensionless radial distance $r\sqrt{n_2}$
(in units of the mean distance between particles of species 2) for a $2D$ 
binary mixture of perfectly permeable particles at $x_1=0.7$. The results 
obtained from numerical simulations with $10^{5}$ Poissonian points are shown
for comparison (dotted lines).} 
\end{figure}

 The results corresponding to a mixture of ideal gases are contained naturally
in the expressions (\ref{e.gl}) and (\ref{e.gij}). In this case, the results  
are exact. For permeable spheres, i.e. $a_{ij}\!=$ $b_{ij}\!=$ 
$a_{ij}^*\!=\!0$, the expected correct result $g_{ij}^*\!=1$ $\forall$ $i,j$
is obtained. Moreover,
\begin{equation} \label{e.gijp}
g_{il,j}(\omega) = \left\{ 
\begin{array}{lr}
1-e^{-n\omega} & (j\ne l) \\
1+(x_l^{-1}-1)e^{-n\omega} & (j=l)
\end{array}
\right..
\end{equation}
 Figure~\ref{f.gide} shows some $cpdf$ corresponding to a binary 
bi-dimensional mixture predicted by Eq. (\ref{e.gijp}) (full lines), which 
are in excellent agreement with the results obtained from numerical 
simulations (dotted lines) based on two sets of uncorrelated points 
created by a random number generator.

\subsection{van der Waals mixture} \label{s.vdw}

 The van der Waals model for a mixture can be obtained by adding to the 
repulsive forces [Eq. (\ref{e.uv_hs})] a generalized K\^ac type attraction 
potential,
\begin{equation} \label{e.uvw}
u_{ilv,j}(\omega) = \left\{ 
\begin{array}{rcl}
 \infty; & & \omega \le a_{ij} \\  
 -\epsilon_{ij} \gamma^* J (\gamma^* \omega); & & \omega > a_{ij}
\end{array}
\right. ,
\end{equation}
\noindent where $\epsilon_{ij}\geq0$ are constants and $J$ is a function that 
satisfy the condition $\int_0^\infty\!J(\omega)d\omega=1$. In the limit
$\gamma^*\!\to\!0$ it becomes, 
\begin{equation} \label{e.phi_ilv} 
\phi_{ilv,l} = \left\{ 
\begin{array}{rcl}
 \infty; & &  v \le b_{il} \\  
        -\epsilon_{il}; & &  v > b_{il}
\end{array}
\right. 
\end{equation}
\noindent and
\begin{equation} \label{e.phi_ijv} 
\phi_{ilv,j \ne l} =  -\epsilon_{ij} ,
\end{equation}
\noindent so that
\begin{equation} \label{e.phi_il} 
\epsilon_{ilv} = \left\{ 
\begin{array}{rcl}
\infty; & &  v \le b_{il} \\  
  - \frac{1}{2}\sum_j\epsilon_{ij}n_j; & & v > b_{il}
\end{array}
\right..
\end{equation}
 The attracting forces modify the chemical potentials and the total 
interaction energy of the gas, so that it results,
\begin{equation} \label{e.a_vdw}
\mu_{il}\!=\!kT\!\left[\!\frac{nb_{il}}{1-nb}\!+\!\ln\!\left(\!\frac{n_{il} 
\lambda^D}{n_i}\frac{n}{1-nb}\!\right)\!\right]\!-\!\sum_j\epsilon_{ji}n_j ,
\end{equation}
\noindent and
\begin{equation} \label{e.a_Uvdw}
\frac{U_{int}} V = -\frac{\epsilon n^2}{2}, 
\end{equation}
\noindent where
\begin{equation} \label{e.b}
b = \sum_{ij}x_ix_j b_{ij},
\end{equation}
\noindent and
\begin{equation} \label{e.epsi}
\epsilon\equiv\sum_{ij}x_ix_j\epsilon_{ij}.
\end{equation}
 Nevertheless, Eqs. (\ref{e.nv}) and (\ref{e.P}) remain unchanged as it 
happens for single-component fluids \cite{paper1}. We then see that using Eq. 
(\ref{e.pgu}) the known relation $P=(nkT)/(1-nb)-\epsilon n^2/2$ is recovered. 
Relations (\ref{e.b}) and (\ref{e.epsi}) are often used as mixing rules which
relate the properties of the pure component models to those of mixtures in 
extended models. They are considered adequate for mixtures of monopolar 
compounds \cite{peng,voros} and vapor-liquid phenomena \cite{han}. 

\subsection{Attractive hard spheres} \label{s.attra}

 A simple model of hard spheres having short-range attractive forces, as
opposite to the long-range van der Waals model, is obtained by adding a 
square-well potential at the edge of a repulsive nucleus,  
\begin{equation} \label{e.uahp}
u_{ilv,j}(\omega) = \left\{ 
\begin{array}{rcr}
\infty; & & \omega \le a_{ij}  \\  
 -\epsilon_{ij}; & &  a_{ij} < \omega < a_{ij} + \xi \\
  0; & & \omega \ge a_{ij} + \xi  
\end{array}
\right.,
\end{equation}
\noindent where $\xi$ is a characteristic attractive volume surrounding the 
$a_{ij}$ volume of the attractive nucleus. For simplicity we adopt a value of 
$\xi$ independent of the considered pair of particles. This potential is used 
often to model the interactions among colloidal particles and protein 
molecules \cite{asherie, zacca, pontoni}, which commonly have attractive 
forces in the very nearest vicinity of their surfaces ($\xi\!\ll\!a_{ij}$). 
With Eq. (\ref{e.uahp}), Eqs. (\ref{e.phi}) and (\ref{e.Eilv}) lead to
\begin{equation} \label{e.pahp}
\phi_{ilv,j}\!=\!\left\{ 
\begin{array}{rcl}
  \infty;& &v \le b_{il} \\  
 -\epsilon_{ij}\!\left(\!\frac{\delta_{lj}}{n_l}\!+\!b_{il}\!+\!\xi\!-
\!v\!\right);& &
               b_{il}\!<\!v\!<\!b_{il}\!+\!\xi \\
  0;& &  v\!\ge\!b_{il}\!+\!\xi
\end{array}
\right.,
\end{equation}
\noindent and
\begin{equation} \label{e.eahp}
\epsilon_{ilv} = \left\{ 
\begin{array}{rcl}
   \infty;& &  v \le b_{il} \\  
-\!\frac{\epsilon_{il}}{2}\!-\!B_i\left(b_{il}\!+\!\xi\!-v\right);& &  
          b_{il} < v < b_{il} + \xi \\
   0;& & v \ge b_{il} + \xi
\end{array}
\right. ,
\end{equation}
\noindent where
\begin{equation} \label{e.Bahs}
B_i = \sum_j\frac{\epsilon_{ij}n_j}{2}. 
\end{equation}
 The state equations that result from the particle and volume conservation 
relations (\ref{e.Nil}) and (\ref{e.Vcons}) are:
\begin{equation} \label{e.muahs}
\mu_{il}= kT\left[\beta\gamma b_{il}+\ln\left(\frac{\beta\gamma n_{il}
\lambda_i^D}{n_i X_{il}}\right)\right]+\epsilon_{i},
\end{equation}
\begin{equation} \label{e.bgahs}
\beta \gamma = \sum_{il}n_{il}\frac{Y_{il}}{X_{il}}, 
\end{equation}
\noindent where we have
\begin{equation}\label{e.Xahs}
X_{il} = e^{-\beta \gamma\xi}+\beta\gamma e^{\beta(\epsilon_{il}/2+B_i\xi)}
\left[\frac{1-e^{-\beta(\gamma+b)\xi}}{\beta(\gamma+B_i)}\right],
\end{equation}
\begin{equation}\label{e.Yahs}
\begin{array}{ll}
Y_{il} = & \left[1+\beta\gamma\left(b_{il}+\xi\right)\right]
e^{-\beta\gamma\xi} \\
 & +\left(\frac\gamma{\gamma+B_i}\right)^2 e^{\beta(\epsilon_{il}/2+B_i\xi)}
\left\{1+\beta (\gamma +B_i)b_{il}\right. \\ 
 & \left.-\left[1+\beta(\gamma+B_i)(b_{il}+\xi)\right]e^{-\beta(\gamma+B_i)\xi}
\right\}. \\
\end{array}
\end{equation}
Besides, the $v$--occupation number distribution is given by
\begin{equation}\label{e.nva}
n_{ilv} =  n_{il} \frac{\beta \gamma}{X_{il}} 
e^{-\beta[\gamma (v-b_{il}) + \epsilon_{ilv}]}.
\end{equation}
 We can also obtain an analytic expression for the interaction energy:
\begin{equation}\label{e.Uahs}
\begin{array}{ll} 
\frac{U_{int}}{V}= & \sum_{il}\frac{\beta\gamma n_{il}}{X_{il}}e^{\beta 
(\epsilon_{il}/2+B_i\xi)} \\
 & \times\left\{-\left(\frac{\epsilon_{il}}{2}+B_i\xi\right)\left(
\frac{1-e^{-\beta(\gamma+B_i)\xi}}{\beta(\gamma+B_i)} \right) 
\right. \\
 & \left.+B_i\left(\frac{1-[1+\beta(\gamma+B_i)\xi]e^{-\beta(\gamma+B_i)\xi}}
{[\beta(\gamma+B_i]^2}\right)\right\}. \\ 
\end{array}
\end{equation}
 It can be interesting to have limiting expressions of these equations 
for small enough values of  $\xi$. Keeping only the zero and first order 
terms, we derive
\begin{equation} \label{e.Xahsx}
X_{il} = 1+\beta\gamma\xi\left(e^{\beta\epsilon_{il}/2}-1\right),
\end{equation}
\begin{equation} \label{e.Yahsx}
Y_{il} = 1+\beta\gamma b_{il}+(\beta\gamma)^2\xi b_{il} 
\left(e^{\beta\epsilon_{il}/2}-1\right).
\end{equation}
 If also the potential wells are the same, $\epsilon_{il}=\epsilon$ 
$\forall(i,l)$, the state equation (\ref{e.bgahs}) reduces to a quadratic 
relation in $\beta \gamma$,
\begin{equation} \label{e.Gahsx}
(\beta\gamma)^2\xi(e^{\beta\epsilon/2}-1)+\beta\gamma = \frac{n}{1-bn},
\end{equation}
\noindent where $b=\sum_{il}x_ix_lb_{il}$. When the thickness $\xi$ or the 
depth $\epsilon$ of the potential well tend to zero, the state equation 
(\ref{e.Gahsx}) recovers the form corresponding to a mixture of pure hard 
spheres. 
The same result is obtained for high enough temperatures, $kT\gg\epsilon$.
 
\subsection{Space distributions}

 The thermodynamic variable $\gamma$ has the function of adjusting the
allotment of space among particles in order to satisfy the mathematical 
requirements of space filling established by (\ref{e.Vcons}). In mixtures of
non-interacting particles the occupation number distributions $n_{ilv}$ 
depends on the individual volume per particle $v$ through the exponential 
$\exp(-\beta\gamma v)$. In this case, the scale factor $\beta\gamma$ is
equivalent to the total number density of particles $n$. Obviously, the 
allotment of space is not homogeneous. Most particles have smaller available 
volumes than the average volume per particle, $n^{-1}$, while the statistical
partition of big volumes is low.\par
\begin{figure}
\scalebox{0.40}{\includegraphics{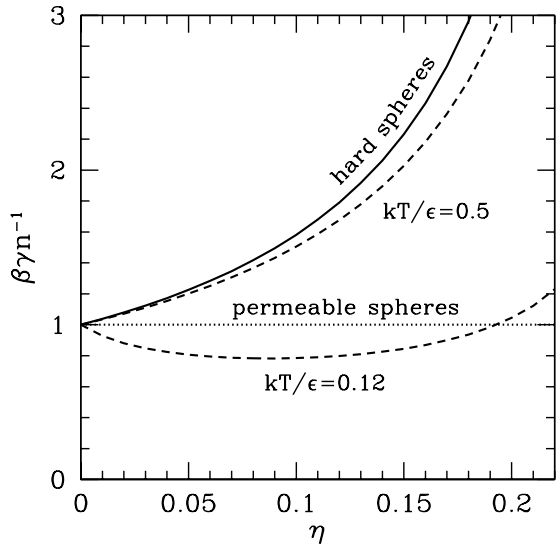}}
\caption{\label{f.gbn} Factor $\beta \gamma n^{-1}$ as a function of the 
packing fraction for $3D$ binary mixtures of spheres with $x_1=0.7$ and 
$d_2/d_1=0.5$. {\em Dotted line}: perfectly permeable spheres; {\em full line}:
hard spheres and van der Waals fluid; {\em dashed lines}: square-well fluid 
with $\xi/\sigma=0.2$ ($\sigma$ is the mean volume of a particle) and reduced
temperatures $kT/\epsilon =0.12$, 0.5. The result for pure hard spheres is 
obtained for $kT/\epsilon\to\infty$.} 
\end{figure}
\begin{figure}
\scalebox{0.40}{\includegraphics{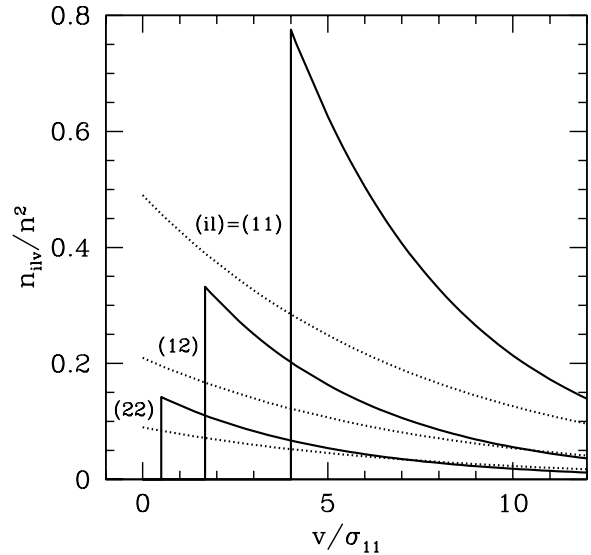}}
\caption{\label{f.nv} Dimensionless double densities $n_{ilv}/n^2$ as a 
function of the available volume per particle (in units of volume $\sigma_{11}$
of a particle of class $i=1$) for a $3D$ binary mixture of hard spheres with 
$\eta=0.1$, $x_1=0.7$ and $d_2/d_1=0.5$ (solid lines). Results corresponding 
to fully penetrable spheres are shown by comparison (dotted lines).} 
\end{figure}
\begin{figure}
\scalebox{0.40}{\includegraphics{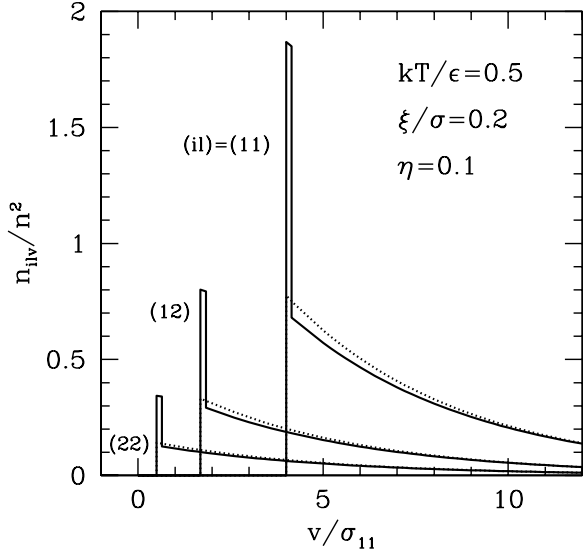}}
\caption{\label{f.nva} Dimensionless double densities $n_{ilv}/n^2$ as a 
function of the available volume per particle (in units of volume $\sigma_{11}$
of a particle of class $i=1$) for a $3D$ binary mixture of attractive hard 
spheres with $\eta=0.1$, $x_1=0.7$, $d_2/d_1=0.5$, $\xi/\sigma=0.2$ and
$kT/\epsilon=0.5$ (solid lines). 
Results for pure hard spheres are shown by comparison (dotted lines).} 
\end{figure}

The particle interactions change the assignment of the volume $v$ for each 
particle and thus they change the occupation numbers $n_{ilv}$, as compared to 
those for ideal gases. Therefore, the interactions also affect the $\gamma$ 
values. In Fig. \ref{f.gbn} we have plotted $\beta \gamma n^{-1}$ as a 
function of the packing fraction $\eta$ for perfectly permeable spheres 
(dotted line), pure hard spheres (full line), and hard spheres with 
square-well potentials (dashed lines). Notice that $\beta \gamma n^{-1}$ is a
measure of $\gamma$ in units of perfect gas values. Figs. \ref{f.nv} and 
\ref{f.nva} show occupation number distributions $n_{ilv}$ for binary mixtures 
of pure and attractive hard spheres, based on Eqs. (\ref{e.nv}) and 
(\ref{e.nva}), respectively. The lower cutoff value for each curve corresponds 
to the smallest volume $v$ permitted by the repulsion between a hard sphere 
and its $nscd$, as given by Eq. (\ref{e.vid}). The square-well potential 
yields the peaks observed in Fig. \ref{f.nva}. 
If we make abstraction of the normalization constants - area under the 
curves representing the $n_{il}$ population - 
the peak height of curves $n_{ilv}$ vs. $v$ in Figs. \ref{f.nv} 
and \ref{f.nva} depend on the abundance of the corresponding pairs $il$. 
In the frame of models of mixtures considered here the distributions $n_{12v}$ 
and $n_{21v}$ are coincident.\par 
In a hard sphere mixture, the repulsions among the particles produce an 
increase of the mean volume per particle. As a consequence, at given 
temperature and density the value of $\gamma$ must increase (full line in 
Fig. \ref{f.gbn}) to warrant the occupation distributions $n_{ilv}$ be 
correctly displaced towards higher values of $v$ (full lines in Fig. 
\ref{f.nv}). 
On the contrary, the attractive forces of finite range in a square-well fluid, 
increase the population of particles with small individual volumes (full
lines in Fig. \ref{f.nva}) and reduce the values of $\gamma$ as compared to 
those of pure hard spheres (dashed lines in Fig. \ref{f.gbn}). These changes
depend on the temperature value. Low temperatures favor the particle
adherence and thus increase the occupation number of low available volume.
\par
The long-range interactions of K\^ac type analized in Sect. \ref{s.vdw} are 
such that the magnitude of the interaction of each particle with the remaining 
medium does not depend on the nearness of its neighbors. Hence, the attractive 
forces have no incidence on the space occupation distribution $n_{ilv}$ 
neither on $\gamma$, both being equivalent to those in a mixture of pure hard 
spheres and hence independent of the temperature.

It is useful to point out that, for ideal and pure hard-sphere mixtures, the 
quantity $\beta \gamma n^{-1}$ coincides with the compressibity factor 
$Z=P/nkT$ since these gases cannot store interaction energy and, therefore, 
$\gamma$ is equivalent to the gas pressure $P$. It is not true for the van der 
Waals and square-well models, because they have a non negligible interaction 
energy, so that $\gamma$ and $P$ differ as described by relation 
(\ref{e.pgu}).

\section{Chemical Equilibrium}  \label{s.chem}

 In this section we extend the theory somewhat to include cases where chemical 
reactions can occur. In this case, the total free energy contains a 
contribution given by the summation of all particle internal energies,
\begin{equation}
E = \sum_iN_iE_i = \sum_{il}\int_0^V\!N_{ilv}E_idv, 
\end{equation}
where $E_i$ is the internal energy of a particle $i$ and the reference zero 
energy is considered the same for all species. If the internal energies 
$E_i$ are degenerate with multiplicity $w_i$, then $W_i$ in Eq. (\ref{e.Wi}) 
becomes,
\begin{equation} \label{e.Wi2}
W_i = N_i!{\prod_{lv}}\frac{(w_idv/V)^{N_{ilv}dv}}{(N_{ilv}dv)!}.
\end{equation}
 The configuration free energy takes thus the form
\begin{equation}
F_{conf} = \sum_{il}\int_0^V\frac{N_{ilv}}\beta\ln\left(\frac{N_{ilv}V}
{w_iN_i}\right)dv.
\end{equation}
 The results obtained in previous sections given by (\ref{e.Nv}, \ref{e.Zilg}, 
\ref{e.Zig}) can be generalized now as,
\begin{equation} \label{e.Nv2}
N_{ilv} = \frac{N_i}{\lambda_i^D}w_i\exp\left[-\beta\left(\gamma v-\mu_{il}+
\epsilon_{ilv}+\epsilon_i+E_i\right)\right],
\end{equation}
\begin{equation} \label{e.Zil2}
Z_{il} = (\beta\gamma)^{-1}w_iq_{il}e^{-\beta(\epsilon_i+E_i)},
\end{equation}
\noindent and
\begin{equation}\label{e.Zi2}
Z_i = (\beta\gamma)^{-1}w_iQ_ie^{-\beta(\epsilon_i+E_i)}.
\end{equation}
 Now, let us consider the reaction
\begin{equation}
aA + bB + ... \longleftrightarrow cC + ...,
\end{equation}
\noindent that involves the species $i=A,B,C,...$ and the stoichiometric 
factors $a,b,c,...$. The equilibrium condition of the reaction is 
\begin{equation}
dF = \sum_i\left(\left.\frac{\partial F}{\partial N_i}\right|_{T,V,N_{j\ne 
i}}\right)dN_i = \sum_i\mu_idN_i = 0,
\end{equation}
\noindent which is subject to the conditions
\begin{equation}
\frac{dN_A} a = \frac{dN_B} b = ... = - \frac{dN_C} c = ...\ .
\end{equation}
 The criterion for the equilibrium is therefore 
\begin{equation}
a\mu_A+b\mu_B+ ... = c\mu_C+ ... \ ,
\end{equation}
\noindent which through Eq. (\ref{e.Nig}) leads to the relationship among 
the species abundances 
\begin{equation}
\left(\frac{N_A\lambda_A^D}{NZ_A}\right)^a\left(\frac{N_B\lambda
_B^D}{NZ_B}\right)^b ... = \left(\frac{N_C\lambda_C^D}{NZ_C}\right)^c
...\ ,
\end{equation}
\noindent that gives the generalized form of the mass-action law.\par
 The results given in this section bring useful tools to study gaseous
mixtures, including their chemical equilibrium. Eq. (\ref{e.Nv2}) sums up the
main feats of the present formulation. It enables to describe populations in
reacting gases at a level of details attained never before, i.e. to evaluate
populations of atoms and molecules in a consistent thermodynamic way, not
only by obtaining the density of each species, but giving as well its 
populations per groups according to the available volume of each particle, 
where this variable is linked directly with the degree of perturbations
experienced by the atoms due to the presence of neighbors. In the preceding
work \cite{paper1} we have foreseen some results that could be obtained with
the present theory. For example, it is possible to show that atoms of a given
species in a gas do not have all the same effective ionization potential 
(electronic energy relative to the energy threshold corresponding to free
electrons), but can they have a variety of values according to the available
volume of particles. The application of the present theory to mixtures of
atomic particles with more realistic interaction potentials than considered
here will be presented in a forthcoming paper.

\section{Conclusions} \label{s.concl}

 We have developed a mathematical framework in terms of a Helmholtz 
free-energy model which aims at describing equilibrium properties of 
$D$-dimensional multi-component fluids. This theory is an extension of a 
former model devoted to one-component fluids and combines free-energy 
minimization methods with space partitions based on the available volume $v$ 
to a particle and the occupation number distributions of this variable. Thus 
we obtain an unified formalism that treats thermodynamics and structure of 
complex fluids simultaneously.\par
 The theory is presently developed for the case of dilute fluids, where 
inter-particle interactions are evaluated from pair potentials. Two closure 
relations (\ref{e.nilninl}) and (\ref{e.vid}) are required to obtain a closed
formalism. The former is associated to the particle number conservation 
(\ref{e.Nil}) and provides the probability that a particle had a specific
species as the closest neighbor. The second closure relation (\ref{e.vid})
specifies the available-volume variable and, therefore, is linked to the 
filling space condition (\ref{e.Vcons}). Consequently, unconventional 
thermodynamic variables were derived by introducing space partitions in the
thermo-statistical analysis of fluids. They are Lagrange's multipliers related
to the specific particle number and volume conservations, which describe thus 
chemical potentials and the pressure, respectively. In addition, we have 
obtained general expressions for the excess chemical potentials and chemical 
equilibrium conditions of non-ideal fluids, as well as several pair 
distribution functions.\par
 The theory provides a straightforward way to calculate thermodynamic 
quantities and structure properties of gas mixtures. As illustrative examples,
we have considered several hard sphere systems for which simple analytical 
expressions were derived. The equations of state obtained for pure hard-sphere 
systems are in good agreement with the scaled-particle and Percus-Yevick 
solutions valid for low-densities. We reckon that this agreement confirms a
posteriori the validity of the several approaches adopted. Besides, well-known 
solutions for van der Waals mixtures were recovered easily by using 
generalized K\^ac potentials. We also have derived closed analytical forms for 
thermodynamic quantities of square-well mixture fluids. These evaluations 
constitute a reference for future studies of short ranged attractive systems.
The influence of repulsive and attractive forces on the space partition among
particles was analyzed. The results thus obtained express a close relationship 
between statistical geometry and statistical mechanics.\par 
 The formalism introduced in the present work provide a useful tool to 
investigate the structure and the effective interactions in chemical complex
fluids. It is particularly well suited to study some astrophysical gases where 
detailed descriptions of species and interactions are required. This occurs, 
in particular, in atmospheres and circumstellar envelopes, where the density 
can be hardly higher than some $\eta=0.01$, so that the approximations for 
low densities given here can be used safely. We consider that present theory 
has other many applications in applied problems of current interest, such as 
studies of equilibrium properties of colloidal and protein solutions.

Among the most important results obtained in this work, we can quote :\par
\begin{itemize}
\item A functional (or series of functionals for mixing) contains the entire
thermodynamical information of the system and leads to a classification of 
particles according to their distance to the closest neighbor through the
assignment of a given volume available to each particle. So, it is thus
possible to distinguish the atoms according to the different degrees of 
perturbations carried by their neighbors. To our knowledge, there is no other
model able to make this distinction up to now. 
\item The current formalism is able to answer to the following inquiries :\par
\begin{itemize}
\item  What is the probability of finding a particle of class $j$ at a 
distance $r$ from another particle of species $i$ whose closest neighbor is of 
class $k$, knowing that $r$, $i$, $j$ and $k$ are arbitrary ?\par
\item Idem as before, but for a neighbor $k$ situated at a given (arbitrary) 
distance from the particle of class $i$ ?\par
\end{itemize}
\item  The chemical equilibrium  of a fluid and the classification of 
particles according to the distance to the respective closest neighbor, are 
done simultaneously and self-consistently with the laws of thermodynamics. 
\item As far as we know, it is the first time that a theory is able to provide 
thermodynamical variables associated with the distribution of the space 
between the particles of different chemical species directly. The variables 
are related to the conventional thermodynamical variables explicitly, such as
pressure, chemical potential, etc.\par
\end{itemize}
 The description level attained by our formalism overrides whatever known
pre-existing model and constitutes an unparalleled tool to represent chemical
mixing at low densities. It also provides conceptual elements on the 
connection between statistical mechanics and spatial statistics, which has
never been presented up to now.\par

\begin{acknowledgments}
We wish to acknowledge useful discussions with Drs. O. Moreschi and A. Santos. 
R.D.R. acknowledges the Carrera del Investigador Cient\'ifico, Consejo de 
Investigaciones Cient\'ificas y T\'ecnicas de la Naci\'on (CONICET), Argentina.
We warmly thank the referees of this paper for their helpful comments 
and suggestions, which helped to improve an earlier version of this paper.
\end{acknowledgments}

\appendix
\section{Thermodynamic identities} \label{s.iden}

 In what follows we derive three thermodynamic identities which enable us to 
derive relations (\ref{e.muil}), (\ref{e.pgu}) and (\ref{e.mui2}). The 
starting point is the equation that gives the total particle population in 
terms of the occupation number $N_{jmv}$,
\begin{equation} \label{e.N}
N = \sum_{jm}\int_0^V\!N_{jmv}dv.
\end{equation}
 If in (\ref{e.N}) we replace $N_{jmv}$ by its expression given by (\ref{e.Nv})
and derivate it with respect to a given population $N_{il}$ by keeping 
constant the temperature, volume and the remaining populations 
($N_{jm\ne il}$), we shall obtain
\begin{equation}  \label{a.der_n}
\begin{array}{l}
\sum_{jm}\int_0^V\!N_{jmv}\!\left[\!v\frac{\partial\gamma}{\partial N_{il}}
- \frac{\partial\mu_{jm}}{\partial N_{il}}\!+\!\frac{\partial\epsilon_{jmv}}
{\partial N_{il}}\!+\!\frac{\partial\epsilon_j}{\partial N_{il}}\!\right] dv 
= \\ 
V\frac{\partial\gamma}{\partial N_{il}}\sum_{jm}N_{jm} 
\frac{\partial\mu_{jm}}{\partial N_{il}} \\ 
+\sum_{jm}\int_0^V\!N_{jmv}\!\left[\!\frac{\partial\epsilon_{jmv}}
{\partial N_{il}}+\frac{\partial\epsilon_j}{\partial N_{il}}\!\right]dv 
= 0.
\end{array}
\end{equation}
\noindent The second line in (\ref{a.der_n}) was obtained using relations
(\ref{e.Nil}) and (\ref{e.Vcons}). We also note that
\begin{equation}
\begin{array}{l}
\sum_{jm}\int_0^V\!N_{jmv}\frac{\partial\epsilon_{jmv}}{\partial N_{il}}dv
= \\ \sum_{jm}\int_0^V\!N_{jmv}\frac{\phi_{jmv,i}}{2V}dv = \epsilon_i,
\end{array}
\end{equation}
\noindent and
\begin{equation} \label{a.a3}
\sum_{jm} \int_0^V N_{jmv} \frac{\partial \epsilon_j}{\partial N_{il}} dv =
\sum_{j} N_j \frac{\partial \epsilon_j}{\partial N_{il}}  =
\frac{\partial U_{int}}{\partial N_{il}} -\epsilon_i \,. 
\end{equation}
 Eqs. (\ref{a.der_n}) to (\ref{a.a3}) lead to the identity
\begin{equation}\label{a.ide}
V \frac{\partial \gamma}{\partial N_{il}} - \sum_{jm} N_{jm} 
\frac{\partial \mu_{jm}}{\partial N_{il}} 
+ \frac{\partial U_{int}}{\partial N_{il}} = 0 \;.
\end{equation}
\noindent Then, Eq. (\ref{e.muil}) can be obtained easily from relations 
(\ref{e.Fequil}) and (\ref{a.ide}). In a similar way, the partial derivative
of Eq. (\ref{e.N}) with respect to the volume produces
\begin{equation} \label{a.der_V}
\begin{array}{l}
\sum_{jm}\!\int_0^V\!N_{jmv}\!\left[v\frac{\partial\gamma}{\partial V}
-\frac{\partial\mu_{jm}}{\partial V}\!+\!\frac{\partial\epsilon_{jmv}}
{\partial V}\!+\!\frac{\partial\epsilon_j}{\partial V}\!\right]dv 
= \\ 
V\frac{\partial\gamma}{\partial V}\!-\!\sum_{jm} N_{jm} 
\frac{\partial\mu_{jm}}{\partial V}\!+\!\frac{\partial U_{int}}{\partial V}\! 
+\!C = 0,
\end{array}
\end{equation}
\noindent with
\begin{equation}
C= \sum_{jm}\!\int_0^V\!N_{jmv}\frac{\partial\epsilon_{jmv}}{\partial V}dv.
\end{equation}
 In Eq. (\ref{a.der_V}) we have ignored the terms with $N_{jm(v=V)}$, since
they become zero at the thermodynamic limit. According to Eq. (\ref{e.Eilv}), 
we have
\begin{equation}
\frac{\partial \epsilon_{jmv}} {\partial V} = - \frac { \epsilon_{jmv} } V 
+ \sum_i \frac{ N_i \partial_V \phi_{jmv,i} } {2V} \,.
\end{equation}
\noindent In general $\partial_V \phi_{jmv,i}=0$, so that
\begin{equation}
C=-\frac 1V \sum_{jm}\int_0^V N_{jmv} \epsilon_{jmv} dv= - \frac{U_{int}}V \;.
\end{equation}
\noindent Hence, from relation (\ref{a.der_V}) we obtain the following 
identity
\begin{equation} \label{a.ide2}
V \frac{\partial \gamma}{\partial V} - \sum_{jm} N_{jm} \frac{\partial 
\mu_{jm}}{\partial V} + \frac{\partial U_{int}}{\partial V} 
-\frac{U_{int}}{V} = 0.
\end{equation}
 Knowing that $P=-\partial F/\partial V$, relation (\ref{e.pgu}) is obtained
straightforwardly from Eqs. (\ref{e.Fequil}) and (\ref{a.ide2}).\par
 The procedure followed to obtain relation (\ref{a.ide}) can also be used 
to obtain the partial derivative of Eq. (\ref{e.N}) with respect to the
population of class $i$ particles ($N_i$) by considering constant the 
temperature, volume and the populations of the remaining species 
($N_{j\ne i}$). We then use
\begin{equation}
\frac{\partial U_{int}}{\partial{N_i}}=\epsilon_i+
\sum_jN_j\partial_{N_i}\epsilon_j,
\end{equation}
\noindent and
\begin{equation}
\frac{\partial\epsilon_{jmv}}{\partial{N_i}}=\frac{\phi_{jmv,i}}{2V} .
\end{equation}
 The third identity that we can obtain is
\begin{equation} \label{a.ide3}
V\!\!\left.\frac{\partial\gamma}{\partial N_{i}}\!\right|_{N_{j\ne i}}\! 
-\!\!\sum_{jm}\!\!N_{jm}\!\! 
\left.\frac{\partial\mu_{jm}}{\partial N_{i}}\right|_{N_{j\ne i}}\!\!+\!\! 
\left.\frac{\partial U_{int}}{\partial N_{i}}\right|_{N_{j\ne i}}\!\!=\!0.
\end{equation}
 Equation (\ref{a.ide3}) enable us to find an expression for the chemical 
potential $\mu_i$ in terms of the thermodynamic potentials $\mu_{jm}$ 
[Eq. (\ref{e.mui2})].


\end{document}